\documentclass[12pt]{article}
\usepackage{amssymb,amsthm,amsfonts}
\usepackage{epsfig}
\usepackage[T1]{fontenc}
\usepackage[latin2]{inputenc}
\usepackage{colordvi}
\def\bea{\begin{eqnarray}}
\def\eea{\end{eqnarray}}
\def\bq{\begin{quote}}
\def\eq{\end{quote}}

\def\gappeq{\mathrel{\rlap
{\raise.5ex\hbox{$>$}}
{\lower.5ex\hbox{$\sim$}}}}
\def\lappeq{\mathrel{\rlap{\raise.5ex\hbox{$<$}}
{\lower.5ex\hbox{$\sim$}}}}

\newcommand{\beq}{\begin{equation}}
\newcommand{\eeq}{\end{equation}}

\def\arg{\mathrm{arg}\,}
\def\varkappa{\omega}

\newcounter{mnotecount}[section]

\begin{document}
\pagestyle{empty}
\begin{flushright}
IFT-2004/31\\
CERN-PH-TH/2004-263\\
{\bf \today}
\end{flushright}
\vspace*{5mm}
\begin{center}

{\LARGE \bf Update on Fermion Mass Models}
\vspace*{.3cm}
{\LARGE \bf with an Anomalous Horizontal $U(1)$ Symmetry}\\
\vspace*{2cm}

{\bf Piotr~H.~Chankowski}$\,^a$, {\bf Kamila~Kowalska}$\,^a$,
{\bf St\'ephane~Lavignac}$\,^{b,}$\footnote{Permanent address:
Service de Physique Th\'eorique, CEA-Saclay, F-91191 Gif-sur-Yvette
Cedex, France.}
and {\bf Stefan~Pokorski}$\,^{a,b}$\\
\end{center}

\noindent
$^a$ Institute of Theoretical Physics, Warsaw University, Ho\.za 69, 00-681, 
Warsaw, Poland\\
$^b$ Theory Division, Physics Department, CERN, CH-1211 
Geneva 23, Switzerland\\

\vspace*{1.5cm} 
\centerline{\bf Abstract} 
\vspace*{5mm}
\noindent
{
We reconsider models of fermion masses and mixings based on a gauge anomalous 
horizontal $U(1)$ symmetry. In the simplest model with a single flavon field
and horizontal charges of the same sign for all Standard Model fields,
only very few charge assignements are allowed when all experimental data,
including neutrino oscillation data, is taken into account. 
We show that a precise description of the observed fermion masses and
mixing angles can easily be obtained by generating sets of the order one
parameters left unconstrained by the $U(1)$ symmetry.
The corresponding Yukawa matrices show several interesting features which 
may be important for flavour changing neutral currents and CP violation 
effects in supersymmetric models.
}
\vspace*{1.0cm}
\date{\today} 


\vspace*{0.2cm}

\vfill\eject

\newpage

\setcounter{page}{1}
\pagestyle{plain}

\section{Introduction}

The origin of the fermion mass and mixing textures is a challenge for 
physics beyond the Standard Model. One promising approach to that problem 
is based on hypothetical horizontal symmetries which are spontaneously 
broken by vacuum expectation values of some ``flavon'' fields $\Phi$. 
Hierarchical patterns in the fermion mass matrices can then be explained 
by the Froggatt-Nielsen mechanism \cite{FRONI}, as due to suppression factors 
$\left(\langle\Phi\rangle/M\right)^n$, where $M$ is the scale of integrated 
out physics and the power $n$ depends on the horizontal group charges of 
the fermion, Higgs and flavon fields.

Fermion mass models based on abelian
\cite{BIWE}--\cite{abelian}
and non-abelian \cite{nonabelian} horizontal symmetries 
have been widely discusssed in the literature. Each 
approach has its own virtues and shortcomings, particularly when realized in 
supersymmetric models, which we consider here. $U(2)$ and $SU(3)$ symmetries 
are quite predictive for fermion mass matrices once the pattern of symmetry 
breaking is specified. However, the quantitative description of fermion 
masses and mixings requires a rather complicated structure of symmetry 
breaking.
The main shortcoming of $U(1)$ horizontal symmetries as models of fermion
masses is the dependence of the quantitative predictions on arbitrary order 
one coefficients, the remnant of the integrated out unknown physics. In 
principle also the arbitrariness in the choice of the abelian charges for 
fermions looks less appealing than the rigid structures of the essentially 
unique choices of $U(2)$ and $SU(3)$ as continuous horizontal non-abelian 
symmetry groups. But that last point is balanced by the fact that the 
breaking of the $U(1)$ symmetry is much simpler.

$U(1)$ gauge group factors are generic in string models. One particularly 
simple and attractive horizontal symmetry is the anomalous $U(1)$ often 
found in heterotic string compactifications. Anomaly cancellation by the 
Green-Schwarz mechanism \cite{GRSC}
is possible only under certain conditions which 
strongly constrain the possible choices of horizontal $U(1)$ charges for 
fermions. It has already been noticed long ago that those constraints are at 
a qualitative level amazingly consistent with the quark masses and mixings
\cite{IBRO}--\cite{BILARA}.
A particularly appealing feature of the anomalous $U(1)$ is that the symmetry 
breaking parameter $\langle\Phi\rangle/M$ can be computed in terms of the 
horizontal charges, and that its value in explicit models turns out to be 
very close to the Cabibbo angle. Furthermore, with positive charges for
matter fields and negative for the flavon field (or vice-versa), this vacuum
is unique.

The purpose of this note is to update the predictions for fermion masses of 
the simplest, string-inspired gauge anomalous $U(1)$ models with a single 
flavon field $\Phi$. We are motivated by several 
factors. One is the recent experimental progress in neutrino physics.
Neutrino masses and mixings can be well described by models based on 
horizontal $U(1)$ symmetries, and it is therefore interesting to check what 
phenomenological constraints, in addition to the Green-Schwarz anomaly 
cancellation conditions, are put on the horizontal charges by the requirement 
that the same gauged $U(1)$ symmetry explains both the quark and lepton 
mass hierarchies. In fact, assuming that all Standard Model fields have 
horizontal charges of the same sign, we find a very limited number of 
possible charge assignments. 

Furthermore, it is interesting to check the success of these models beyond
the qualitative level. In this paper, we perform a complete numerical fit
of a few representative models to fermion masses and mixings, with the
inclusion of complex order one coefficients and with attention paid to
the $\tan\beta$ dependence.

Finally, in supersymmetric models there are new sources of flavour and
CP violation induced by virtual sfermion exchanges. Sfermion mass matrices
depend both on the pattern of supersymmetry breaking and on the rotations
to the super-CKM basis which are determined by the fermion mass matrices. 
Thus, it is useful to have an explicit set of successful fermion mass models,
defined by a set of horizontal charges and complex order one coefficients. 
The fact that only a small number of charge assignments are allowed makes 
it possible to test the phenomenological implications of an anomalous gauge 
horizontal symmetry. In a forthcoming paper, we shall use this set of models 
to study flavour changing neutral currents (FCNCs) in supersymmetry breaking 
scenarios with dominant $D$-term breaking.

The paper is organized as follows. In the next section, we review the
basic properties of fermion mass models based on an anomalous $U(1)$
and list the phenomenologically allowed charge assignements.  
In section 3 we generate sets of complex order one coefficients giving
a precise description of fermion masses and mixings, and study the
features of the associated Yukawa matrices relevant for FCNC and CP
violating processes. Finally, we present our conclusions in section 4.

\section{Constraints on $U(1)_X$ charges}
\label{sec:constraints}

The models we consider in this paper are extensions of the
($R$-parity conserving) MSSM
with a horizontal gauge abelian symmetry $U(1)_X$ and a chiral superfield
$\hat\Phi$ with $X$-charge normalized to $-1$, whose vacuum expectation
value breaks the horizontal symmetry.
The $X$-charges of the MSSM superfields $\hat Q_A$, $\hat U^c_A$, 
$\hat D^c_A$, $\hat L_A$, $\hat E^c_A$ ($A=1,2,3$), $\hat H_u$ and $\hat H_d$ 
are denoted by $q_A$, $\bar u_A$, $\bar d_A$, $l_A$, $\bar e_A$, $h_u$ and 
$h_d$, respectively. The model also contains three right-handed neutrino 
superfields $\hat N^c_A$ with charges $\bar n_A$, which are needed to 
generate the neutrino masses via the seesaw mechanism \cite{seesaw}.
No additional matter
charged under the SM gauge group is assumed, although the forthcoming
analysis would not be altered by the presence of vector-like matter under
both the SM gauge group and $U(1)_X$. The Yukawa couplings 
of quarks and leptons are generated via the Froggatt-Nielsen mechanism 
\cite{FRONI}, from nonrenormalizable superpotential terms of the form:
\begin{equation}
C_u^{AB}\, \hat U^c_A \hat Q_B  \hat H_u 
\left({\hat\Phi\over M}\right)^{\bar u_A+q_B+h_u} ~.
\label{eqn:nonren}
\end{equation}
These terms arise upon integrating out heavy vector-like (super)fields,
the so-called Froggatt-Nielsen fields, whose characteristic mass scale is 
$M$. In string compactifications, the r\^ole of the Froggatt-Nielsen fields 
is played by massive string modes, and $M$ is identified with 
the Planck scale (although vector-like fields may also be present among the 
massless string modes). After breaking of the horizontal symmetry by the VEV 
of the scalar component $\phi$ of the superfield $\hat\Phi$, one obtains 
effective Yukawa couplings suppressed by powers of the small parameter 
$\epsilon\equiv\langle\phi\rangle/M$:
\begin{equation}
\mathbf{Y}_u^{AB} =C_u^{AB} ~\epsilon^{\bar u_A+q_B+h_u}~ .
\label{eqn:Yu}
\end{equation}
The factors $C_u^{AB}$ are not constrained by the horizontal symmetry and
are assumed to be of order one. Then the hierarchy of Yukawa 
couplings is determined, up to these unknown factors, by the charges of 
the MSSM fields. Since holomorphicity of the superpotential forbids 
nonrenormalizable terms with a negative power of the superfield $\hat\Phi$,
one has $\mathbf{Y}_u^{AB}=0$ if \footnote{This conclusion can 
be evaded if the Froggatt-Nielsen fields, instead of being vector-like under 
both the Standard Model gauge group and the $U(1)_X$ group as usually assumed, 
are chiral under $U(1)_X$. In such a case effective operators carrying a 
negative $X$-charge can be induced in the low-energy effective theory. 
Moreover, even if the Froggatt-Nielsen fields are vector-like under 
$U(1)_X$, effective operators $\hat L_A\hat L_B\hat H_u\hat H_u$ with a 
negative $X$-charge may be induced by the seesaw mechanism, since right-handed 
neutrino superfields are chiral under $U(1)_X$ (see e.g. the model of Ref. 
\cite{IRLARA}).} $\bar u_A+q_B+h_u<0$. Similarly, $\mathbf{Y}_u^{AB}=0$ if
$\bar u_A+q_B+h_u$ is not an integer. 

Within the above assumptions, one can show that the horizontal abelian 
symmetry has to be anomalous \cite{BILARA} (see also Refs. \cite{IBRO,BIRA}). 
Indeed, one has the following relations:
\begin{equation}
\det M_u \det M_d \sim v^3_u v^3_d~\epsilon^{C_3+ 3(h_u+h_d)}~,
\end{equation}
\begin{equation}
\det M_d/\det M_e ~\sim ~
\epsilon^{-{1\over2}(C_1+C_2-{8\over3}C_3)+h_u+h_d}~,
\label{eq:Md_over_Me}     
\end{equation}
where $C_3\equiv\sum_A(2q_A+\bar u_A+\bar d_A)$, 
$C_2\equiv\sum_A(3q_A+l_A)+h_u+h_d$ and 
$C_1\equiv{1\over3}\sum_A(q_A+8\bar u_A+2\bar d_A+3l_A+6\bar e_A)+h_u+h_d$ 
are the coefficients of the mixed $SU(3)_C$-$SU(3)_c$-$U(1)_X$, 
$SU(2)_L$-$SU(2)_L$-$U(1)_X$ and $U(1)_Y$-$U(1)_Y$-$U(1)_X$ anomalies, 
respectively, and $v_{u,d}$ are the vacuum expectation values of the two 
Higgs doublets. If all anomaly coefficients were vanishing, the hierarchy 
among quark masses would require a large, positive value of $h_u+h_d$ 
($h_u+h_d=6 - 8$ for $\epsilon = \lambda$, where $\lambda$
is the sine of the Cabibbo angle, depending on the value of $\tan\beta$),
while the relation (\ref{eq:Md_over_Me}) would require a much smaller
value ($h_u+h_d=1$ or $2$ for $\epsilon = \lambda$). 
Therefore, the horizontal symmetry has to be anomalous if it is to explain 
the observed fermion mass hierarchy.

This fact provides the main motivation for considering a gauged horizontal 
symmetry. It is well-known that abelian gauge anomalies can be compensated
for by the Green-Schwarz mechanism \cite{GRSC}, as is common in 
four-dimensional heterotic string compactifications\footnote{Anomalous 
$U(1)$'s are also common in four-dimensional open string compactifications. 
However, while heterotic string compactifications contain at most one 
anomalous abelian gauge group factor, open string compactifications
may contain several anomalous $U(1)$'s whose anomalies are compensated for by 
a generalized Green-Schwarz mechanism \cite{SA}.
In this case, the Green-Schwarz anomaly conditions are less constraining than 
in the heterotic case, and the scale of breaking of the anomalous $U(1)$'s 
depends on twisted moduli vevs, which are fixed by unknown nonperturbative
physics. For these reasons, we prefer to consider the case of an anomalous 
$U(1)$ of the heterotic type, with its anomalies canceled by the universal 
Green-Schwarz mechanism.}.
This requires that the following relations between anomaly coefficients
be satisfied:
\begin{equation}
{C_1\over k_1} = {C_2\over k_2} = {C_3\over k_3} = {C_X\over k_X}
={\mbox{Tr} X\over12}~, \phantom{aaa} C_{XXY} = 0 ~, 
\label{eq:GS_cond}
\end{equation}
where $C_X$ is the coefficient of the cubic $U(1)_X$ anomaly, $\mbox{Tr}X$ 
is the coefficient of the mixed $U(1)_X$-gravitational anomaly, and $k_a$ 
is the Kac-Moody level of the gauge group $G_a$, which depends on the 
compactification. The Kac-Moody level of a non-abelian gauge group is an 
integer, while the Kac-Moody level of an abelian gauge group can be fractional.
In the following, we shall assume $k_2=k_3$, as is the case in most of (if 
not all of) the heterotic string compactifications constructed so far. Since
the $U(1)_X$-$U(1)_X$-$U(1)_Y$ anomaly cannot be compensated for by the 
Green-Schwarz mechanism, its coefficient $C_{XXY}$ has to vanish by itself.

A nice feature of the Green-Schwarz mechanism is that the value of the 
Weinberg angle at the string scale is determined by the ratio of the anomaly 
coefficients $C_1$ and $C_2$ \cite{IB}. Indeed, upon combining the anomaly 
conditions with the gauge coupling unification relation 
$k_1g^2_1=k_2g^2_2=\cdots=g^2_{\rm string}$ valid at the string
scale, one obtains:
\begin{equation}
\sin\theta^2_W (M_{\rm string}) = {C_2\over C_1+ C_2}~ .
\end{equation}
The canonical value of the Weinberg angle at the GUT scale
($\sin^2\theta_W=3/8$) is obtained for $C_1/C_2=5/3$, or equivalently 
for the normalization $k_1/k_2=5/3$, which is well known from string model 
builders. As observed in Refs.~\cite{IBRO,BIRA,NIR}, this
normalization (together with $k_2=k_3$) leads through
Eq. (\ref{eq:Md_over_Me}) to
${m_d m_s m_b\over m_e m_\mu m_\tau}\sim\epsilon^{h_u+h_d}$, which given
the uncertainty due to the order one coefficients is consistent with
data\footnote{Assuming approximate bottom-tau Yukawa coupling
unification at the GUT scale, this mass ratio is between
0.75 and 0.9 for the central values of the fermion masses in the
$\tan\beta$ range $5-60$, which favours $h_u + h_d = 0$.}
for $h_u + h_d = 0, \pm 1$ or even $\pm 2$.
Such values of $h_u+h_d$ are suitable for electroweak symmetry breaking if
the $\mu$-term is generated from the Giudice-Masiero mechanism \cite{GIMA}.
In the following, we shall fix the Weinberg
angle at its canonical GUT value and impose $C_1/C_2=5/3$.

Another nice feature of an anomalous $U(1)_X$ symmetry is that its breaking 
scale, hence the small expansion parameter $\epsilon$, is determined 
by the value of the gauge coupling at the unification scale and by the
anomaly coefficient $C_2$ \cite{IRLARA}. Indeed, due to the fact that 
$\mbox{Tr}X\neq0$, a Fayet-Iliopoulos term is generated at the one-loop 
level \cite{DISEWI,ATDISE}:
\begin{equation}
\xi^2 ={g^2_{\rm string}\over192\pi^2}~ \mbox{Tr}(X M^2)\ ,
\end{equation}
where $g_{\rm string}$ is the string coupling and $M$ the reduced Planck 
mass. Provided that $X_\Phi\mbox{Tr}X <0$, this triggers a nonzero VEV
of the scalar component $\phi$ of the superfield $\hat\Phi$, which sets 
the value of the $\epsilon$ parameter to 
\cite{IRLARA}:
\begin{equation}
\epsilon =\sqrt{{g^2_{\rm string}\over192 \pi^2}~\mbox{Tr}X} =
\sqrt{{\alpha_2\over4\pi} C_2}\ ,
\label{eq:epsilon}
\end{equation}
where we have used the relations $k_2g^2_2=g^2_{\rm string}$ and 
$C_2/k_2=\mbox{Tr}X/12$ (recall that $X_\Phi=-1$). As we shall see later, 
in realistic models the value of $\epsilon$ is very close to the 
Cabibbo angle, $\lambda \simeq 0.22$.

We now look for solutions to the anomaly constraints. In addition, we 
require the absence of kinetic mixing between the anomalous $U(1)$ and the 
hypercharge, i.e. $\mbox{Tr}(XY)=0$. As noticed in Ref. \cite{NEWR}, this 
condition also prevents the generation of a large one-loop Fayet-Iliopoulos 
term for the hypercharge \cite{DIGIU} in the scenario in which the dominant
contribution to soft scalar masses comes from the anomalous $D$-term. Such
a Fayet-Iliopoulos term could have induced charge and colour breaking minima. 
We look for models with the $X$-charges of all quark, lepton and Higgs
superfields positive or zero, which is sufficient to ensure the uniqueness
of the vacuum that breaks the anomalous $U(1)_X$ (see Ref. \cite{IRLA} for
a detailed discussion of flat directions).

The relevant anomaly coefficients read:
\begin{eqnarray}
C_1&=&{1\over3}\sum_A(q_A+8\bar u_A+2\bar d_A+3l_A+6\bar e_A)+h_u+h_d~, \\
C_2&=&\sum_A(3q_A+l_A)+h_u+h_d~, \\
C_3&=&\sum_A(2q_A+\bar u_A+\bar d_A) ~, \\
C_{XXY}&=&\sum_A(q^2_A-2\bar u^2_A+\bar d^2_A-l^2_A+\bar e^2_A)+h^2_u-h^2_d~,\\
C_X&=&{2\over3}\sum_A(6q^3_A+3\bar u^3_A+3\bar d^3_A+2l^3_A+\bar e^3_A+n^3_A)
      +h^3_u+h^3_d-1+C^\prime_X ~, \\
\mbox{Tr}X&=&\sum_A(6q_A+3\bar u_A+3\bar d_A+2l_A+\bar e_A+\bar n_A)
             +2(h_u+h_d)-1+C^\prime_g~ ,
\end{eqnarray}
where $C^\prime_X$ and $C^\prime_g$ stand for the contributions of additional 
SM singlets or vector-like representations charged under $U(1)_X$
(which may be charged under some hidden gauge group) to the cubic $U(1)_X$
anomaly and to the mixed gravitational anomaly, respectively.
Finally, $\mbox{Tr}(XY)$ reads:
\begin{equation}
\mbox{Tr}(XY)=2\sum_A(q_A-2\bar u_A+\bar d_A-l_A+\bar e_A)+2(h_u-h_d)~. 
\label{eq:Tr_XY}
\end{equation}

The constraints\footnote{We assume that $C^\prime_X$, $C^\prime_g$
and the Kac-Moody level $k_X$ are such that the Green-Schwarz conditions
are satisfied for the cubic $U(1)_X$ anomaly and for the mixed
gravitational anomaly.}:
\begin{equation}
C_2=C_3~, \phantom{aaa}C_2={3\over5}~C_1~, \phantom{aaa} C_{XXY} =0
\phantom{aa}\mbox{and}\phantom{aa}\mbox{Tr}(XY) =0~,
\end{equation}
can be rewritten as:
\begin{eqnarray}
&&\sum_A(\bar u_A-q_A)-{1\over2}h_u=0~, \label{eq:constraint_1} \\
&&\sum_A(\bar d_A-l_A)-{1\over2}h_u-h_d=0~, \label{eq:constraint_2} \\
&&\sum_A(\bar u_A-\bar e_A)-h_u=0~, \label{eq:constraint_3} \\
&&\sum_A(q^2_A-2\bar u^2_A+\bar d^2_A-l^2_A+\bar e^2_A)+h^2_u-h^2_d=0 ~.
\label{eq:constraint_4}
\end{eqnarray}

We look for solutions to Eqs. (\ref{eq:constraint_1})-(\ref{eq:constraint_4})
which successfully reproduce the observed quark and lepton mass and mixing 
hierarchies with all matter charges non-negative. Since the Yukawa 
couplings are generated at the scale of breaking of the anomalous $U(1)_X$ 
(which for definiteness we identify with the scale at which gauge couplings
unify in the MSSM, $M_{GUT} \simeq 2\times10^{16}$ GeV), we must consider
the masses and mixings renormalized also at this scale. In the quark sector,
the renormalization of the CKM matrix affects, to a very good approximation, 
only the parameter $A$ of the Wolfenstein parametrization, while $\lambda$, 
$\rho$ and $\eta$ almost do not evolve with energy 
\cite{OLPO,BABU,BABEOH,CHPO2}. The scale dependence of the quark mass ratios 
$m_d/m_s$ and $m_u/m_c$ is also very weak. 
Numerically for $\tan\beta$ values in the range $3.5 - 50$ we find:
\begin{eqnarray}
{m_s(\mu)\over m_b(\mu)}=\chi(\mu)~{m_s(M_Z)\over m_b(M_Z)}\ ,
\phantom{aaa}
{m_c(\mu)\over m_t(\mu)}=\chi^3(\mu)~{m_c(M_Z)\over m_t(M_Z)}\ ,
\end{eqnarray}
with 
$\chi(M_{GUT})\simeq (0.75-0.9)$. This leads to the following approximate 
hierarchies of quark mixing angles and mass ratios close to the GUT 
scale, expressed in powers of the Cabibbo angle $\lambda\simeq0.22$:
\begin{eqnarray}
|V_{us}|\simeq|V_{cd}|\simeq\lambda~, \phantom{aa} 
|V_{ub}|\sim\lambda^4~, \phantom{aa} 
|V_{cb}|\sim\lambda^2~,\phantom{aa}
|V_{td}|\sim\lambda^4 - \lambda^3~,\phantom{aa}  
|V_{ts}|\sim\lambda^2~,\label{eqn:orders1}
\end{eqnarray}
\begin{eqnarray}
{m_d\over m_s}\sim\lambda^2~, \phantom{aa} 
{m_s\over m_b}\sim{1\over2}\lambda^2~,\phantom{aa} 
{m_u\over m_c}\sim\lambda^4~, \phantom{aa}  
{m_c\over m_t}\sim\lambda^4~.\label{eqn:orders2}
\end{eqnarray}
For light quarks we have used here the central values of the PDG
estimates \cite{PDG04}. The dependence on $\tan\beta$ of the GUT scale 
values of these mass ratios and CKM elements  is weak in the 
considered range of $\tan\beta$ values.

In the lepton sector, the charged lepton mass hierarchy is:
\begin{eqnarray}
{m_e\over m_\mu}\sim{1\over2}\lambda^3~, \phantom{aa}  
{m_\mu\over m_\tau}\sim\lambda^2~.
\end{eqnarray}
If one assumes a hierarchical neutrino mass spectrum, i.e. 
$m_{\nu_3}\approx\sqrt{\Delta m^2_{\rm atm}}$ and 
$m_{\nu_2}\approx\sqrt{\Delta m^2_{\rm sol}}$, the scale dependence of 
lepton mixing angles and neutrino masses is very weak \cite{CHPO,CHPO2}, 
and one can take, at the scale where right-handed neutrinos decouple:
\begin{eqnarray}
|U_{\alpha A}|\sim1~,\phantom{aa} (\alpha,A)\neq(e,3)~,\phantom{aaa}   
|U_{e3}| < \lambda~, \phantom{aaa} 
{m_{\nu_2}\over m_{\nu_3}}\sim\lambda~,\phantom{aaa} 
{m_{\nu_1}\over m_{\nu_2}} <1~.
\end{eqnarray}

Eqs. (\ref{eq:constraint_1}) to (\ref{eq:constraint_4}), together with the 
assumption of positive $X$-charges for the MSSM superfields, are very 
restrictive and only a limited number of charge assignments compatible with 
these constraints give a satisfactory account of the observed mass and mixing 
hierarchies. Let us first consider the quark sector. The large top quark 
Yukawa coupling is accounted for by choosing $q_3=\bar u_3=h_u=0$. 
Furthermore, the quark mass ratios and mixing angles are given by the simple 
expressions:
\begin{eqnarray}
&{m_{u_A}\over m_{u_B}}\sim\epsilon^{(q_A-q_B) + (\bar u_A-\bar u_B)}~, 
\phantom{aaa} 
{m_{d_A}\over m_{d_B}}\sim\epsilon^{(q_A-q_B) + (\bar d_A-\bar d_B)}~, 
\phantom{aaa} 
{m_t\over m_b}\sim\tan\beta~\epsilon^{-(\bar d_3+h_d)}~, &\nonumber\\
&V_{AB}\sim\epsilon^{|q_A-q_B|}~.&\label{eqn:orders3}
\end{eqnarray}
The symbol $\sim$ reminds us that the above relations contain unknown factors 
of order one, so that the actual values of the mass ratios and mixing angles 
may slightly depart from the naive ``power counting''. With this remark in 
mind, and assuming $\epsilon\simeq\lambda$, we find that the CKM matrix is 
correctly reproduced by $q_A=(3,2,0)$, $q_A=(4,3,0)$ or by $q_A=(4,2,0)$. In 
each case, the naive ``power counting'' disagrees by at most one power of 
$\lambda$ with the measured value of one or two CKM angles: $|V_{ub}|$ for 
$q_A=(3,2,0)$, $|V_{cb}|$ and $|V_{ts}|$ for $q_A=(4,3,0)$, $|V_{us}|$ and 
$|V_{cd}|$ for $q_A=(4,2,0)$. In the lepton sector, the experimental data 
on neutrino oscillations strongly constrain the $l_A$ (for a recent review,
see e.g. Ref. \cite{ALFE}). In the absence of 
holomorphic zeroes in the Dirac and Majorana mass matrices, the seesaw 
mechanism yields an effective light neutrino mass matrix of the form:
\begin{equation}
\mathbf{M}_\nu\sim{v^2_u\epsilon^{2(l_3+h_u)}\over M_R} 
\left(\matrix{
\epsilon^{2(l_1-l_3)} & \epsilon^{(l_1-l_3)+(l_2-l_3)} & \epsilon^{(l_1-l_3)}  
\cr
\epsilon^{(l_1-l_3)+(l_2-l_3)} & \epsilon^{2(l_2-l_3)} & \epsilon^{(l_2-l_3)}  
\cr
\epsilon^{(l_1-l_3)} & \epsilon^{(l_2-l_3)} & 1 }\right) ~,
\end{equation}
where $M_R$ is the scale of right-handed neutrino masses.
Such a mass matrix can easily reproduce the hierarchical neutrino mass 
spectrum. 
In order to reproduce both the large atmospheric mixing angle and the 
hierarchy between the atmospheric and solar mass scales, one must choose 
$l_2=l_3$ and allow for a mild tuning between the order one entries in the 
lower right $2\times2$ submatrix of $\mathbf{M}_\nu$, of order 
$\sqrt{\Delta m^2_{\rm sol}/\Delta m^2_{atm}} \approx 0.2$.
Due to this tuning, the solar mixing angle $\theta_{12}$ comes out
``large'' provided that $l_1-l_3=1$ or $2$ (see Appendix~\ref{app:PMNS}
for details). This in turn implies 
that $|U_{e3}|\sim\epsilon^{l_1-l_3}$ should be rather large, and even close 
to its present experimental limit in the case $l_1-l_3=1$.

For a given choice of the $q_A$ and the $l_A$ dictated by the CKM and
PMNS mixing matrices, the remaining $X$-charges of the model are
constrained both by the quark and charged lepton 
masses and by Eqs. (\ref{eq:constraint_1}) to (\ref{eq:constraint_4}).
We list below the solutions for the $X$-charge assignments satisfying the 
constraints (\ref{eq:constraint_1}) to (\ref{eq:constraint_4}) for which the 
predictions obtained from naive power counting are in reasonably good 
agreement with experimental data on fermion masses and mixings, when 
extrapolated to the GUT scale:
\begin{eqnarray}
&\mbox{\bf 1:}& q_A=\bar u_A=\bar e_A=(3,2,0),   \phantom{a} 
                    \bar d_A=l_A=(l_1,l_3,l_3),  \phantom{a} h_u=h_d=0,
                    \nonumber \\
                     & & [l_1-l_3=1\ \mbox{or}\ 2;\ l_3=0,1, 2\ \mbox{or}\ 3]  
                    \label{eq:sol1} \\
&\mbox{\bf 2:}& q_A=\bar u_A=\bar e_A=(3,2,0), \phantom{a} \bar d_A=(2,1,0), 
                    \phantom{a} l_A=(1,0,0),  \phantom{a} h_u=0, \phantom{a} 
                    h_d=2, \hskip 1cm
                    \label{eq:sol2} \\
&\mbox{\bf 3:}& q_A=\bar u_A=\bar e_A=(3,2,0), \phantom{a} \bar d_A=(1,1,0), 
                    \phantom{a} l_A=(1,0,0),  \phantom{a} h_u=0, \phantom{a} 
                    h_d=1, \hskip 1cm
                    \label{eq:sol3} \\
&\mbox{\bf 4:}& q_A=\bar u_A=\bar e_A=(3,2,0), \phantom{a} \bar d_A=(2,1,0), 
                    \phantom{a} l_A=(2,0,0),  \phantom{a} h_u=0, \phantom{a} 
                    h_d = 1, \hskip 1cm
                    \label{eq:sol4} \\
&\mbox{\bf 5:}& q_A=\bar u_A=e_A=(4,2,0), \phantom{a} 
                    \bar d_A=l_A=(l_3+1,l_3,l_3), \phantom{a} h_u=h_d=0, 
                    \nonumber \\
                     & & [l_3=0,1, 2\ \mbox{or}\ 3]
                    \label{eq:sol5} \\
&\mbox{\bf 6:}& q_A=\bar u_A=\bar e_A=(4,2,0), \phantom{a} \bar d_A=(1,1,0), 
                    \phantom{a} l_A=(1,0,0),\phantom{a} h_u=0, \phantom{a} 
                    h_d=1. \hskip 1cm
                    \label{eq:sol6}
\end{eqnarray}
Several of these solutions can be found in the literature. The constraints 
(\ref{eq:constraint_1}) to (\ref{eq:constraint_4}) are automatically satisfied 
if $h_u=h_d=0$ and the horizontal symmetry commutes with $SU(5)$, as 
already noticed in Ref. \cite{NEWR}. This is the case for solutions {\bf 1}
and {\bf 5}. We were not able to find solutions with $h_u = h_d = 0$ that
are not compatible with the $SU(5)$ symmetry. In all solutions, the naive
predictions slightly depart from the observed values for some quantities,
and one has to rely on the effect of the unconstrained order one parameters
to correct them. In particular, solutions {\bf 1} to {\bf 4} naively
predict $m_u/m_c\sim\lambda^2$ and $|V_{ub}|\sim\lambda^3$, while
solutions {\bf 5} and {\bf 6} predict $|V_{us}|\sim|V_{cd}|\sim\lambda^2$,
but give the correct $m_u/m_c$ ratio. In addition, solution {\bf 1}
with $l_1-l_3=1$ and solutions {\bf 2}, {\bf 3} yield 
$m_e/m_\mu\sim\lambda^2$; solution {\bf 1} with $l_1-l_3=2$ and solution
{\bf 5} yield $m_d/m_s\sim\lambda^3$, and solution {\bf 3} yields
$m_d/m_s\sim\lambda$. Finally, $l_1-l_3=1$ is preferred over $l_1-l_3=2$
by the solar mixing angle \cite{ALFEMA}.

Solutions {\bf 1} and {\bf 5} with $l_3=2$ and solution 2 predict
a low value of $\tan\beta$, $\tan\beta\lesssim15$;
solutions {\bf 1} and {\bf 5} with $l_3=1$ and solutions {\bf 3},
{\bf 4} and {\bf 6} require $\tan\beta\sim(15-50)$; and solutions 
{\bf 1} and {\bf 5} with $l_3=0$ require $\tan\beta\gtrsim50$.
The range of $\tan \beta$ values compatible with a given solution
is rather broad due to the effect of the order one coefficients.

Using Eq. (\ref{eq:epsilon}), one can compute the predicted value
of the expansion parameter $\epsilon$. Assuming 
$\alpha_2(M_{\rm string})=\alpha_U={1\over24}$, one obtains
(the quoted values scale as $(\frac{\alpha_U}{1/24})^{1/2}$):
\begin{eqnarray}
& \epsilon=0.23, ~0.25, ~0.27 & \mbox{for solution 1 with $l_1-l_3=1$
                                      and $l_3 = 0, 1, 2$}, \\
& \epsilon=0.24, ~0.26, ~0.28 & \mbox{for solution 1 with $l_1-l_3=2$
                                      and $l_3 = 0, 1, 2$},  \\
& \epsilon=0.24 & \mbox{for solution 2, 3 and 4},  \\
& \epsilon=0.25, ~0.27, ~0.29 & \mbox{for solution 5 with $l_3=0,1,2$}, \\
& \epsilon=0.26 & \mbox{for solution 6}.
\end{eqnarray}
As already stressed in Ref. \cite{IRLARA}, it is a remarkable success of 
flavour models based on an anomalous $U(1)$ that the predicted value of 
$\epsilon$ comes out so close to the Cabibbo angle. Note that the charge
assignment $q_A = \bar u_A = \bar e_A = (2,1,0)$,
$l_A = \bar d_A = (1,0,0) + l_3$, sometimes considered in the literature,
predicts $\epsilon = (0.18 - 0.23)$, which is too large for this charge
assignment to be compatible with the observed fermion masses and mixings.

\section{Precision description of fermion masses and mixings}

The fact that only a small number of horizontal charge assignments
are phenomenologically acceptable, together with the interesting
 theoretical aspects discussed in the previous section, makes 
the simplest anomalous $U(1)$ models for fermion masses worth further,
more quantitative study. Such a study may also be useful for studying
various aspects of the supersymmetric flavour problem and CP violation
with hierarchical fermion mass matrices. 

The charge assignments {\bf 1}-{\bf 6} in Eqs.~(\ref{eq:sol1})-(\ref{eq:sol6}) 
predict the hierarchical structure of the quark and charged lepton
Yukawa couplings up to order one factors $C^{AB}_{u,d,e}$, which should
be viewed as fixed by some unknown physics which has been integrated out.
The freedom in these factors can be used to obtain a precise description
of the fermion masses and mixings\footnote{For a quantitative study
of the predictions in the neutrino sector, we refer to Ref. \cite{ALFEMA},
which however only considered the choice $l_1-l_3=1$ (``semi-anarchical''
case). We argued, on the basis of the analytical formulae given
in appendix \ref{app:PMNS}, that the choice $l_1-l_3=2$ also leads to a 
satisfactory description of neutrino masses and mixings. In the following,
we shall not discuss the neutrino sector again and shall focus on the
charged fermion sector.}; it is the purpose of this section to find such sets
of coefficients $C^{AB}_{u,d,e}$ and to discuss their properties.

Before doing so, let us notice that, among the acceptable $U(1)_X$
charge assignments, two (namely charge assignments {\bf 1} and {\bf 5})
are consistent with $SU(5) \times U(1)_X$ symmetry, and can therefore
be reconciled with Grand Unification of elementary forces. Moreover, 
those are the only acceptable charge assignments for $h_u = h_d = 0$,
for which, as mentioned earlier, there is some (although not strong)
phenomenological preference. Charge assignments {\bf 3} and {\bf 6},
although not compatible with $SU(5)$ symmetry, are interesting
in the context of the supersymmetric flavour problem because the
right-handed down and strange quark superfields have the same horizontal 
charge, which in some supersymmetry breaking scenarios may suppress
the squark contribution to kaon mixing.

If one interprets the fact that the charge assignments
{\bf 1} and {\bf 5} are compatible with $SU(5) \times U(1)_X$ symmetry
as the manifestation of an underlying Grand Unified Theory, one should
impose the following (GUT-scale) constraints on the order one coefficients:
$C^{AB}_u = C^{BA}_u$ and $C^{AB}_d = C^{BA}_e$. The second constraint
leads to the well-known $SU(5)$ relations $m_e/m_\mu = m_d/m_s$ and
$m_\mu/m_\tau = m_s/m_b$, which are in gross disagreement with the
measured fermion masses and must be corrected \cite{GEJA}.
This can be done through the contribution of renormalizable \cite{GEJA}
or non-renormalizable \cite{ELGA} operators to the Yukawa matrices.
Following Ref.~\cite{ALFEMA_SU5}, we shall introduce
an additional $U(1)_X$ singlet superfield $\hat \Sigma$ transforming
as a $\bf{75}$ of $SU(5)$, which has non-renormalizable couplings
to fermions of the form
$\bar{\mathbf{5}}\, \mathbf{10}\, \bar H\, \hat \Sigma / M$. The Yukawa 
couplings of the down-type quarks and charged leptons then arise from the 
two $SU(5) \times U(1)_X$ invariant superpotential terms:
\begin{eqnarray}
W = \left(
\bar{\mathbf{5}}^A C_1^{AB} \mathbf{10}^B \bar H +
\frac{\hat \Sigma}{M}\, \bar{\mathbf{5}}^A C_2^{AB} 
\mathbf{10}^B \bar H \right)
\left( \frac{\hat \Phi}{M} \right)^{\bar d_A+q_B+h_d} ,
\end{eqnarray}
which, after the scalar components of $\hat\Phi$ and $\hat\Sigma$ acquire VEVs,
lead to:
\begin{eqnarray}
\mathbf{Y}^{AB}_d = \left(C_1^{AB} + \kappa\, C_2^{AB} \right) 
\epsilon^{\bar d_A+q_B+h_d} , \nonumber\\
\mathbf{Y}^{AB}_e = 
\left(C_1^{BA} - 3 \kappa\, C_2^{BA} \right) \epsilon^{\bar d_B+q_A+h_d} ,
\label{eqn:YdYe}
\end{eqnarray}
where $\kappa \equiv \langle\Sigma\rangle/M$. In our numerical fits,
we take $\kappa=0.3$, which makes it easy to account for the difference
between down-type quark and charged letpon masses.

\begin{figure}
\begin{center}
\includegraphics*[height=9cm]{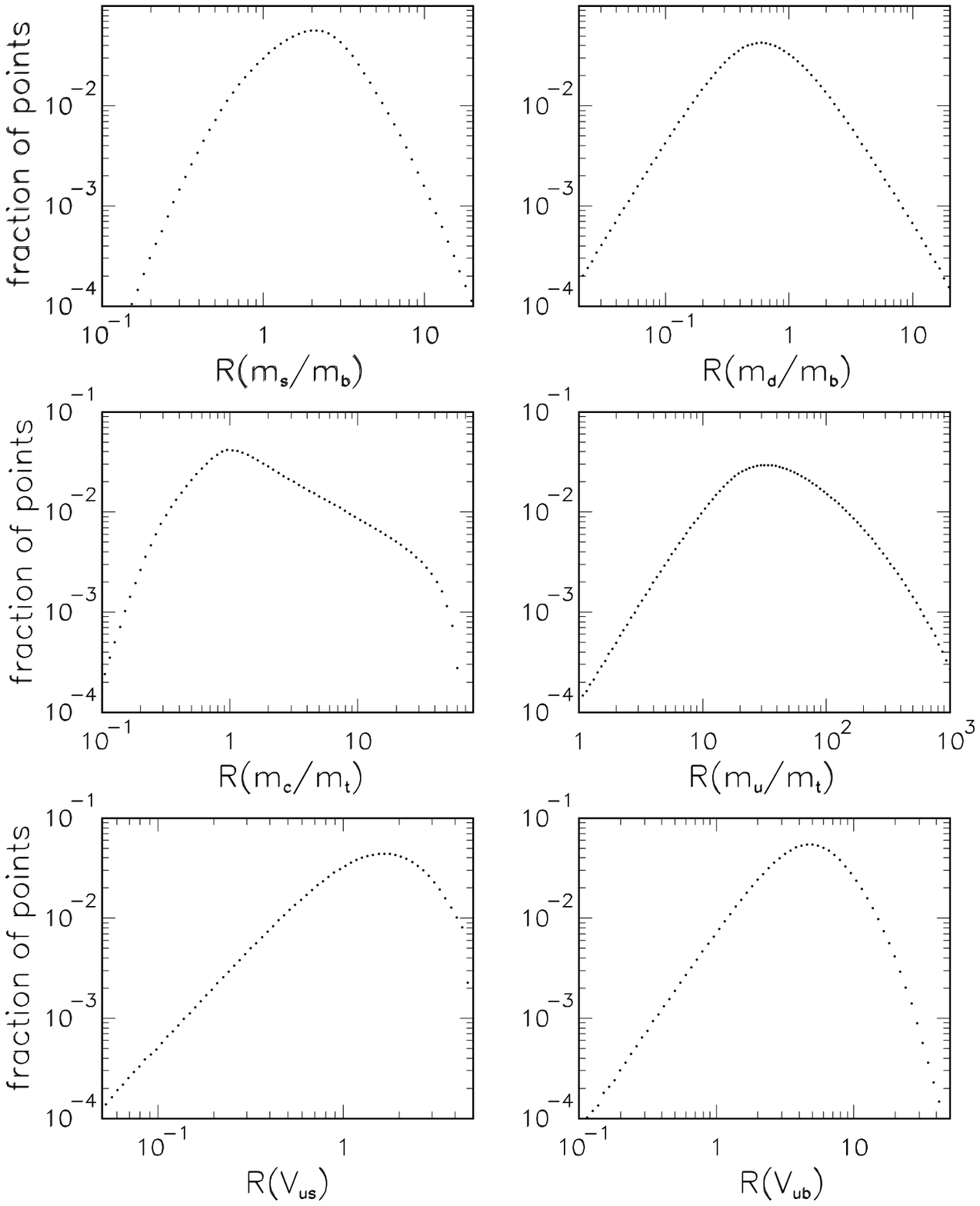}
\hskip 1cm
\includegraphics*[height=9cm]{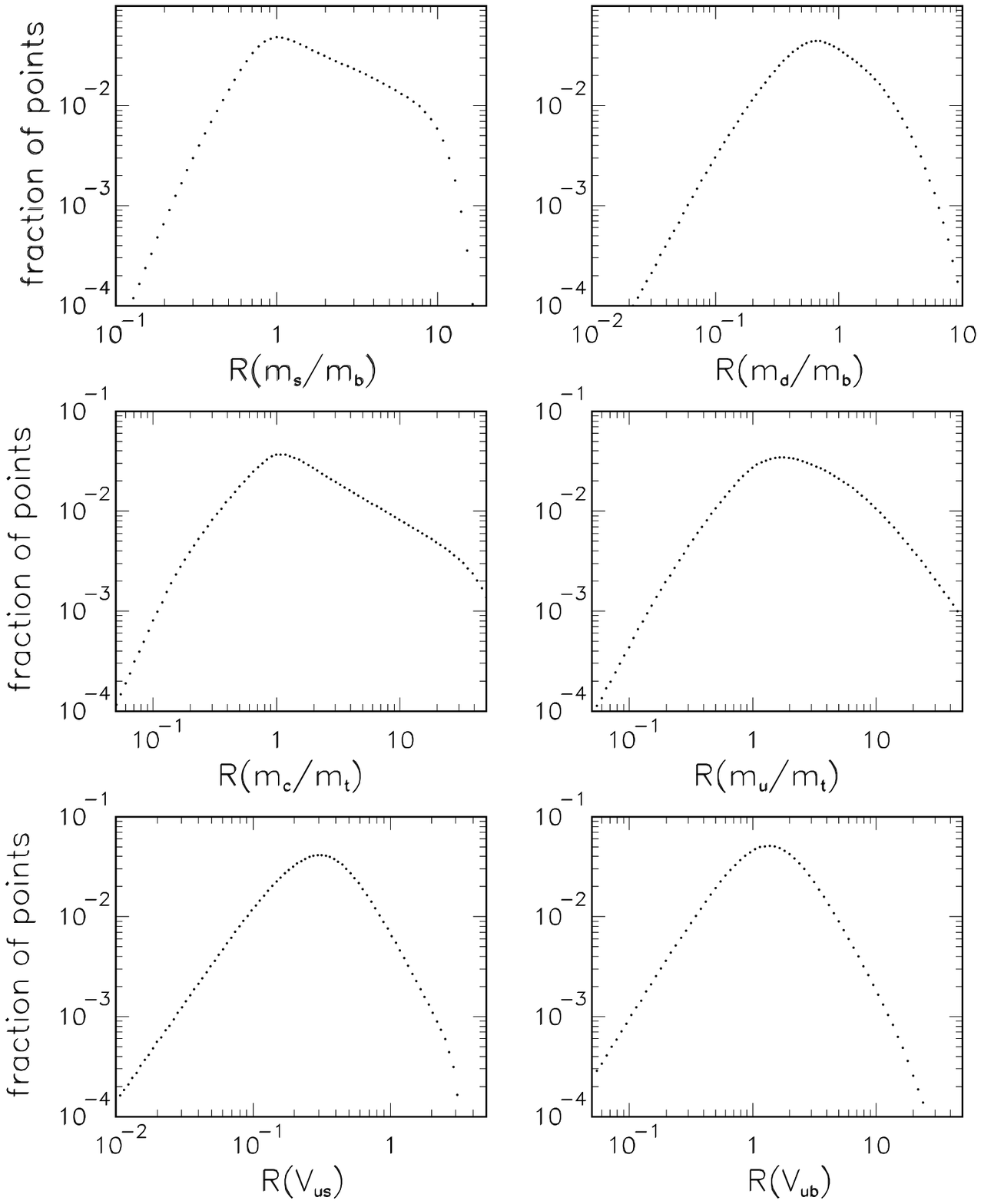}
\caption{Histograms of $R(m_s/m_b)$, $R(m_d/m_b)$, 
$R(m_c/m_t)$, $R(m_u/m_t)$, $R(V_{us})$ and $R(V_{ub})$, where 
$R(Q)\equiv Q(M_{GUT})_{\rm predicted}/Q(M_{GUT})_{\rm evolved}$,
for charge assignment {\bf 1}
with $l_A=\bar d_A=(4,2,2)$ and $\tan\beta=15$
{\it (left set of panels)} and for charge assignment {\bf 6}
with $\tan\beta=45$
{\it (right set of panels)}.}
\label{fig:histogram}
\end{center}
\end{figure}

\begin{figure}
\begin{center}
\includegraphics*[height=6cm]{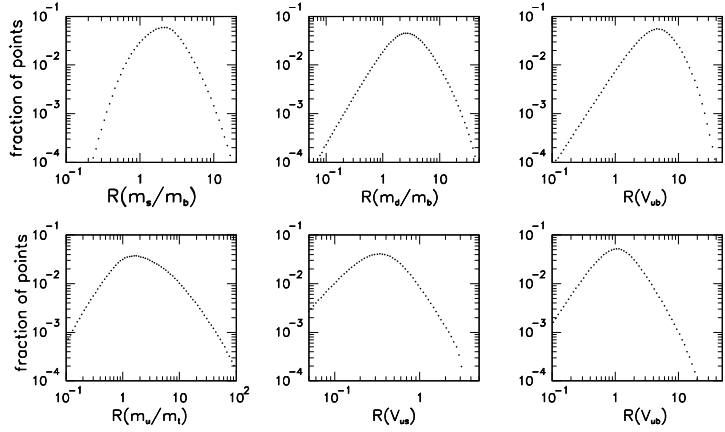}
\caption{Histograms of $R(m_s/m_b)$, $R(m_d/m_b)$ and $R(V_{ub})$
for charge assignment {\bf 1}
with $l_A=\bar d_A=(3,2,2)$ and $\tan\beta=15$ 
{\it (upper set of panels)}, and histograms of $R(m_u/m_t)$,
$R(V_{us})$ and $R(V_{ub})$ for charge assignment {\bf 5}
with $\tan\beta=15$ {\it (lower set of panels)}.}
\label{fig:histogram2}
\end{center}
\end{figure}

In order to test in a quantitative way the ability of the $U(1)_X$
symmetry to describe the fermion masses and mixings, we look for sets
of $24$ complex coefficients $C^{AB}_u = C^{BA}_u$, $C^{AB}_1$ and
$C^{AB}_2$ that reproduce the $9$ quark and charged lepton masses
and the $4$ parameters of the CKM matrix. It is also interesting to
check how well the various charge assignments reproduce the observed
values of the fermion masses and mixings for randomly generated
coefficients.
This is illustrated in Figs.~\ref{fig:histogram} and \ref{fig:histogram2}
for charge assignments {\bf 1} (with two different choices for
$l_A = \bar d_A$), {\bf 5} and {\bf 6}. For randomly generated sets of
coefficients \{$C^{AB}_{u,1,2}$\} with arbitrary phases and moduli
in the range $0.3-3$, each histogram shows the relative number of
sets of coefficients leading to a given value of the ratio
$R(Q)\equiv Q(M_{GUT})_{\rm predicted}/Q(M_{GUT})_{\rm evolved}$
for some particular mass ratio or mixing angle $Q$.
The positions of the peaks are nicely consistent with the naive
``power counting'' of Eq. (\ref{eqn:orders3}) and with the qualitative
discussion of the previous section. Since all charge assignments
{\bf 1}$-${\bf 6} share similar qualitative features, we choose model
{\bf 1} for a more detailed analysis.

These histograms also show that, as expected on the basis of the
qualitative arguments of the previous section, it is easy to find
sets of order one coefficients \{$C^{AB}_{u,1,2}$\} giving a precision
description of fermion masses and mixings. In fact, since the number
of real parameters by far exceeds the number of observables, there are
infinitely many different sets of the coefficients $C^{AB}_{u,1,2}$
that satisfy the physical requirements.
Instead of searching for them by random generation, it is 
much more efficient to use the equations from Appendix \ref{app:CKM} as 
constraints on the generation procedure.

In the qualitative discussion of the previous section, we have used the 
values of the observables at $M_{GUT}$, after evolving them with the RGE from 
their measured low-energy values. For precision fits, this procedure is 
inconvenient because the proper inclusion of experimental errors in the 
renormalization group running is somewhat troublesome. Therefore, we proceed 
top-down. For a given set of complex coefficients chosen with arbitrary phases 
and with their moduli in the range $0.3 - 3$, we calculate the CKM 
matrix entries and the eigenvalues of the Yukawa matrices at the GUT scale 
and evolve them with the renormalization group equations down to the scale 
$M_Z$ \cite{BABU,BABEOH,CHPO2}. The comparison with experimental data is then 
done at $M_Z$, and for a set of coefficients to be acceptable, we require 
that the predicted values of the CKM elements fall in the following 
ranges:
\begin{eqnarray}
0.217<\left|\mathbf{V}_{us}\right|<0.231\ ,  \nonumber\\
0.035<\left|\mathbf{V}_{cb}\right|<0.048\ ,  \nonumber\\
0.065<\left|{\mathbf{V}_{ub}\over\mathbf{V}_{cb}}\right|<0.105\ ,  \nonumber\\
-73^0 <{\rm arg}(\mathbf{V}_{ub})< - 53^o\ .
\end{eqnarray}
The range of ``experimental'' values of the eigenvalues of the Yukawa matrices
at $M_Z$ is obtained by starting with the following set of mass values:
\begin{eqnarray}
&& m_u(2~{\rm GeV})=(2.75\pm2.50)~{\rm MeV}~,\phantom{aaa}
   m_d(2~{\rm GeV})=(6\pm4)~{\rm MeV}, \nonumber\\
&& m_c(2~{\rm GeV})=(1.25\pm0.20)~{\rm GeV}~, \phantom{aaa}
   m_s(2~{\rm GeV})=(105\pm50)~{\rm MeV}~,  \nonumber\\
&& m_b(m_b)=(4.25\pm0.30)~{\rm GeV}~,\phantom{aaaaaa}
   m_t=(178\pm10)~{\rm GeV}~,  \\
&& m_e=(0.511\pm0.025)~{\rm MeV}, \phantom{a}
   m_\mu=(105.6\pm0.5)~{\rm MeV}, \phantom{a}
   m_\tau=(1.778\pm0.1)~{\rm GeV}~.\phantom{aa}  \nonumber
\end{eqnarray}
For quark masses, these ranges are twice as broad as the PDG estimates.
For lepton masses we assigned arbitrary uncertainties, since the actual
experimental errors are much smaller than the uncertainty in the RG
evolution due to e.g. the errors on the gauge couplings
or quark Yukawa couplings, or to the fact that we are only using
two-loop RG equations.
The Yukawa couplings at the $M_Z$ scale are obtained by evolving these masses 
up using the 2-loop QCD RG equations (and taking into account the uncertainty 
in the strong coupling: $\alpha_s(M_Z)=0.1185\pm0.005$).
The top quark Yukawa at the $M_Z$ scale is adjusted iteratively so that
its RG trajectory passes through the right value at the $m_t$ scale. 
In this way we establish the allowed ranges of the Yukawa couplings at the 
$M_Z$ scale. 

\begin{figure}
\begin{center}
\includegraphics*[height=4cm]{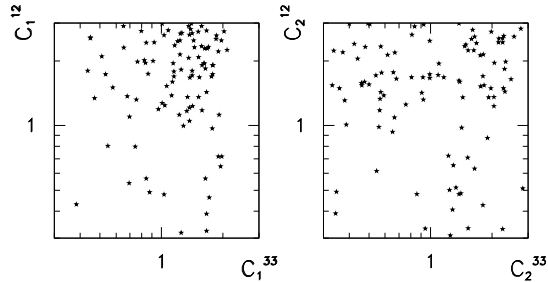}
\caption{Scatter plots of $|C^{12}_1|$ versus $|C^{33}_1|$ {\it (left panel)}
and $|C^{12}_2|$ versus $|C^{33}_2|$ {\it (right panel)} taken from 100
sets of coefficients \{$C^{AB}_{u,1,2}$\} reproducing the observed
masses of quarks and charged leptons as well as the CKM matrix, for
charge assignment {\bf 1} with $l_A = \bar d_A = (4,2,2)$ and
$\tan \beta = 15$.}
\label{fig:cc}
\end{center}
\end{figure}

Once the procedure is specified, one can generate sets of coefficients
\{$C^{AB}_{u,1,2}$\} satisfying the above physical requirement.
For illustration, we show in Fig.~\ref{fig:cc} scatter plots of the
moduli of some coefficients for charge assignment {\bf 1}. One does not
observe any particular correlations;
of course, correlations may (and presumably do) exist in more
dimensional coefficient space.

\begin{figure}
\begin{center}
\includegraphics*[height=8cm]{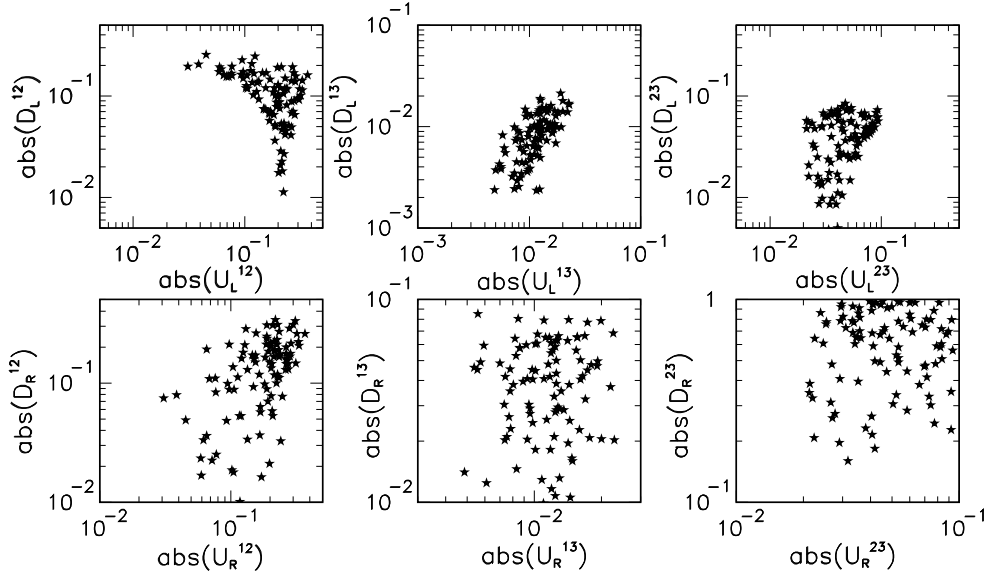}
\caption{Scatter plots of moduli of selected entries of the rotation 
matrices $U_L$, $D_L$, $U_R$ and $D_R$ for charge assignment {\bf 1} 
with $l_A=\bar d_A=(4,2,2)$ and $\tan \beta = 15$.}
\label{fig:rotations}
\end{center}
\end{figure}

\begin{figure}
\begin{center}
\includegraphics*[height=7.5cm]{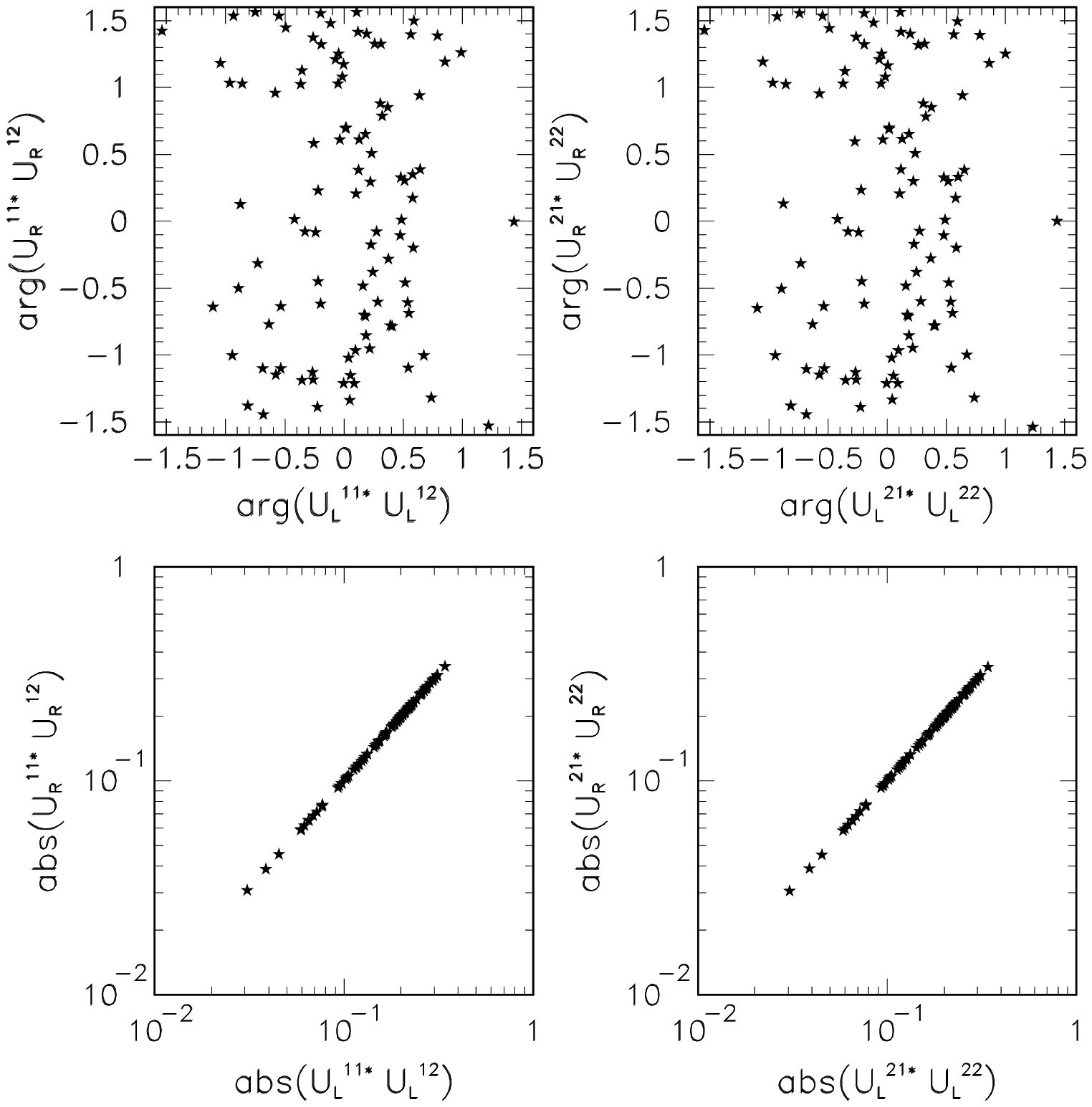}
\hskip 0.5cm
\includegraphics*[height=7.5cm]{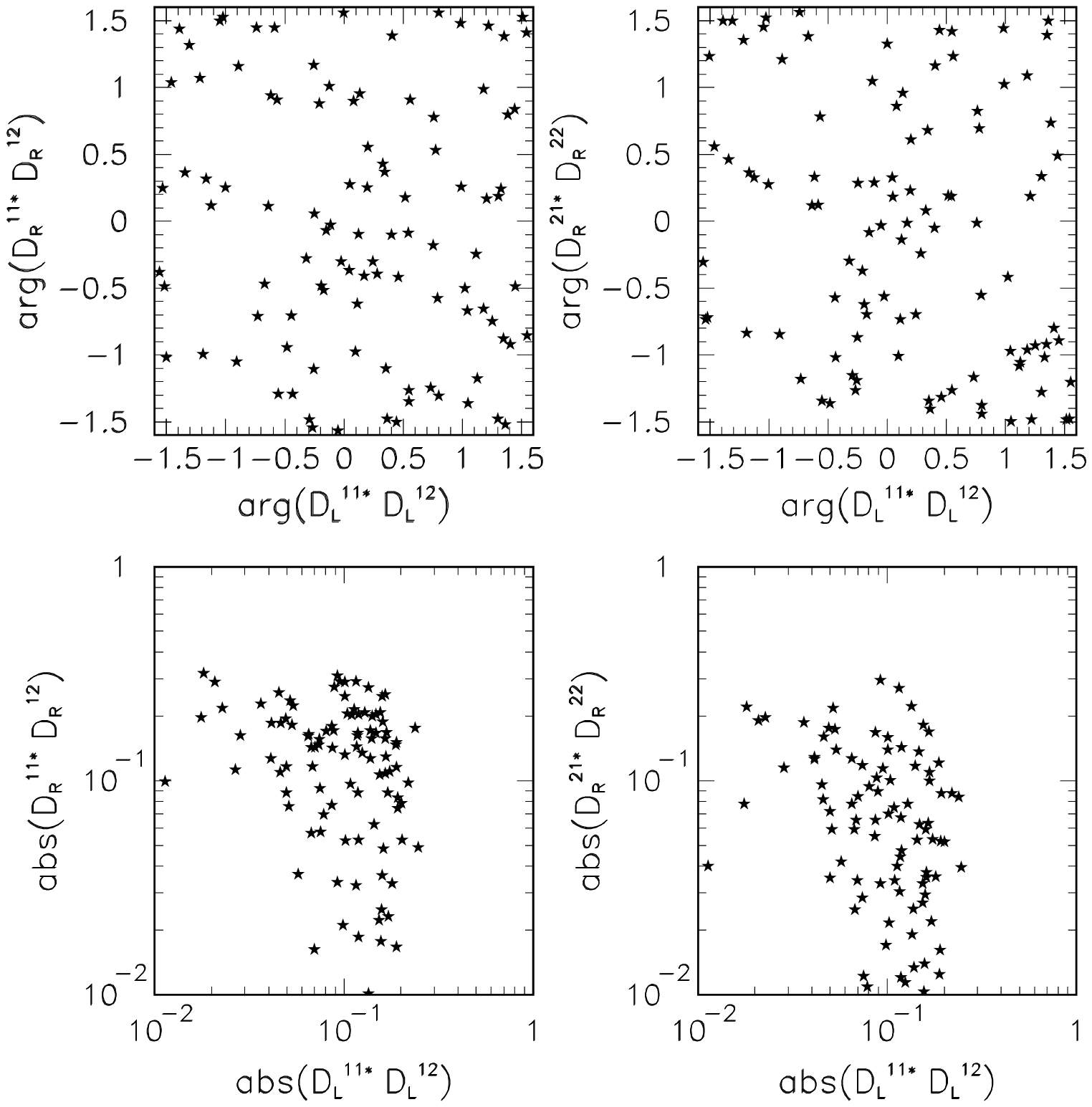}
\caption{Scatter plots of the moduli and phases of the combinations
of the rotation matrices $U_{L,R}$ and $D_{L,R}$ relevant for 
$D^0$-$\bar D^0$ mixing {\it (left set of panels)} and for $K^0$-$\bar K^0$
mixing {\it (right set of panels)} for charge assignment {\bf 1} with 
$l_A=\bar d_A=(4,2,2)$ and $\tan\beta=15$.}
\label{fig:rotations_effects1}
\end{center}
\end{figure}

\begin{figure}
\begin{center}
\includegraphics*[height=8cm]{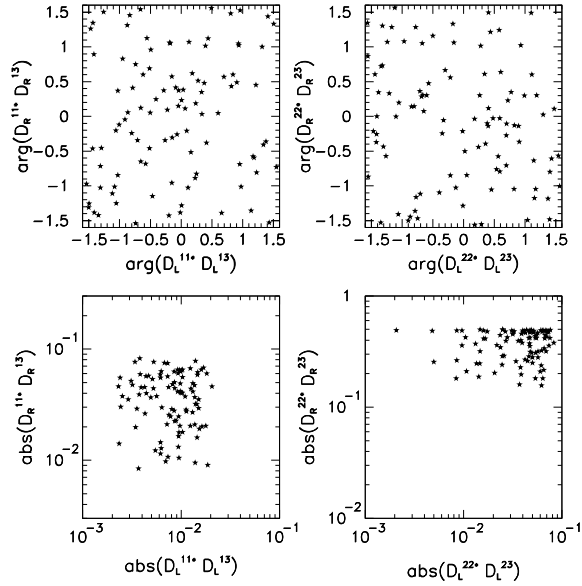}
\caption{Scatter plots of the moduli and phases of the combinations
of the rotation matrices $U_{L,R}$ and $D_{L,R}$ relevant for 
$B_d^0$-$\bar B_d^0$ mixing {\it (left set of panels)} and for
$B_s^0$-$\bar B_s^0$ mixing and $b \rightarrow s$ transitions
{\it (right set of panels)} for charge 
assignment {\bf 1} with $l_A=\bar d_A=(4,2,2)$ and $\tan\beta=15$.}
\label{fig:rotations_effects2}
\end{center}
\end{figure}

The important physical features of the model are shown in 
Figs.~\ref{fig:rotations}, \ref{fig:rotations_effects1} and
\ref{fig:rotations_effects2}, again for charge assignment {\bf 1} with
$l_A = \bar d_A = (4,2,2)$ and $\tan \beta = 15$ taken as an example.
In Fig.~\ref{fig:rotations} we plot the moduli of selected elements
of the rotation matrices
$\mathbf{U}_{L,R}$ and $\mathbf{D}_{L,R}$ defined by
\begin{equation}
U^\dagger_R \mathbf{Y}_u U_L\ =\ 
{\rm Diag}\, (y_u,y_c,y_t)\ , \qquad
D^\dagger_R \mathbf{Y}_d D_L\ =\ 
{\rm Diag}\, (y_d,y_s,y_b)\ .
\end{equation}
The phases of these matrices are chosen in such a way that the Yukawa
matrix eigenvalues are real and that the CKM matrix $V = U_L^\dagger D_L$
is compatible with the standard PDG parametrization. We observe that
$|U^{12}_L|$ and $|D^{12}_L|$ are generically both large, i.e. both
the up and the down quark sectors contribute significantly to the Cabibbo
angle. Similarly, for right-handed quarks,  $|U^{12}_R|$ and $|D^{12}_R|$
are generically both large. Finally, the phases of the rotation matrices
$\mathbf{U}_{L,R}$ and $\mathbf{D}_{L,R}$ (not shown in the plot)
are random and often large.

These features are important for the discussion of FCNC processes and
CP violating effects in supersymmetric models. Indeed, as mentioned in
the introduction, the squark mass matrices depend on the rotations
to the super-CKM basis, which are determined by the fermion mass
models. In order to estimate the effects of these rotations on flavour
and CP violating processes, we assume that, in the ``flavour'' basis
in which the quark superfields have well-defined horizontal charges,
the squark mass matrices are diagonal with non-vanishing entries, e.g.
\begin{equation}
\tilde M^{d2}_{LL}\ =\ \left(\matrix{\tilde m^2_1 & 0 & 0 \cr
                                     0 & \tilde m^2_2 & 0 \cr
                                     0 & 0 & \tilde m^2_3}\right) .
\end{equation}
Then, in the super-CKM basis, the off-diagonal entries of the squark
mass matrix $\left. \tilde M^{d2}_{LL}
\right|_{SCKM} = D^\dagger_L \tilde M^{d2}_{LL} D_L$ are given by:
\begin{eqnarray}
(\tilde M^{d2}_{LL})^{12} & = &
D^{11 \star}_L D^{12}_L (\tilde m^2_1 - \tilde m^2_3)
+ D^{21 \star}_L D^{22}_L (\tilde m^2_2 - \tilde m^2_3)  \\
& \simeq & D^{11 \star}_L D^{12}_L (\tilde m^2_1 - \tilde m^2_2)\ ,  \\
(\tilde M^{d2}_{LL})^{23} & = &
D^{12 \star}_L D^{13}_L (\tilde m^2_1 - \tilde m^2_3)
+ D^{22 \star}_L D^{23}_L (\tilde m^2_2 - \tilde m^2_3)  \\
& \simeq & D^{22 \star}_L D^{23}_L (\tilde m^2_2 - \tilde m^2_3)\ ,  \\
(\tilde M^{d2}_{LL})^{13} & = &
D^{11 \star}_L D^{13}_L (\tilde m^2_1 - \tilde m^2_3)
+ D^{21 \star}_L D^{23}_L (\tilde m^2_2 - \tilde m^2_3)\ ,
\end{eqnarray}
where we have used the fact that $D^{AB}_L \sim \epsilon^{|q_A - q_B|}$
to identify the dominant contribution to $(\tilde M^{d2}_{LL})^{12}$
and $(\tilde M^{d2}_{LL})^{23}$.
Similar formulae hold for the off-diagonal entries of
$\tilde M^{u2}_{LL}$ and $\tilde M^{u2}_{RR}$, with $D_L$ replaced by
$U_L$ and $U_R$, respectively. For $\tilde M^{d2}_{RR}$ one has, for
charge assignments {\bf 1} and {\bf 5}:
\begin{eqnarray}
(\tilde M^{d2}_{RR})^{12} & = &
D^{11 \star}_R D^{12}_R (\tilde m^2_1 - \tilde m^2_3)
+ D^{21 \star}_R D^{22}_R (\tilde m^2_2 - \tilde m^2_3)\ ,  \\
(\tilde M^{d2}_{RR})^{23} & = &
D^{12 \star}_R D^{13}_R (\tilde m^2_1 - \tilde m^2_3)
+ D^{22 \star}_R D^{23} (\tilde m^2_2 - \tilde m^2_3)  \\
& \simeq & D^{22 \star}_R D^{23}_R (\tilde m^2_2 - \tilde m^2_3)\ ,  \\
(\tilde M^{d2}_{RR})^{13} & = &
D^{11 \star}_R D^{13}_R (\tilde m^2_1 - \tilde m^2_3)
+ D^{21 \star}_R D^{23}_R (\tilde m^2_2 - \tilde m^2_3)\ .
\end{eqnarray}
Figs.~\ref{fig:rotations_effects1} and \ref{fig:rotations_effects2}
show the moduli and phases of the combinations of the rotation matrices
$U_{L,R}$ and $D_{L,R}$ relevant for $D^0$-$\bar D^0$, $K^0$-$\bar K^0$,
$B^0_d$-$\bar B^0_d$ and $B^0_s$-$\bar B^0_s$ mixing.
We see that the considered hierarchical models of fermion masses may lead to
large mixings in the $(1,2)$ up squark sector
(the linear correlation seen in the two lower left panels of 
Fig.~\ref{fig:rotations_effects1} is due to the symmetric form
of the matrix $\mathbf{Y}_u$, which implies that 
$U_R = U^\star_L P$, where $P$ is a diagonal matrix of phases)
and to a large mixing in the $\tilde b_R$-$\tilde s_R$ sector.
The order of magnitude of these mixings is consistent with the naive
``power counting'':
$U^{AB}_R \sim U^{AB}_L \sim D^{AB}_L \sim \epsilon^{|q_A-q_B|}$
($U^{AB}_R \sim U^{AB}_L$ is due to the fact that $q_A = u_A$ in
all phenomenologically acceptable charge assignments)
and $D^{AB}_R \sim \epsilon^{|d_A-d_B|}$.
As for the phases in the off-diagonal entries of the squark mass
matrices, one can hardly see any specific pattern in the plots of
Figs.~\ref{fig:rotations_effects1} and \ref{fig:rotations_effects2}.
As one could have anticipated, these phases are essentially random.

This may lead to significant -- or even excessive -- contributions
to flavour changing neutral currents and CP violating processes,
especially in the kaon sector \cite{GAGAMASI,CIU}.
Significant supersymmetric contributions may also be expected
in $D^0$-$\bar D^0$ mixing (with a potentially large CP-violating phase,
see also Ref. \cite{NIRA}), as well as in $B^0_s$-$\bar B^0_s$ mixing
and in $b \rightarrow s$ transitions \cite{CIFRAMASI}. In particular,
large supersymmetric contributions to CP asymmetries in e.g.
$B^0_d \rightarrow \Phi K_S$ or $ B^0_s \rightarrow J/\Psi \Phi$ are
possible.

\section{Conclusions}

We have updated and put at the quantitative precision level the predictions 
for fermion masses and mixings in the simple Froggatt-Nielsen model based on 
a string-inspired gauge anomalous $U(1)$ symmetry spontaneously broken
by a single flavon field, with all matter charges of the same sign.
The inclusion of the neutrino oscillation data adds phenomenological
constraints on the charge assignment, in addition to the Green-Schwarz anomaly
cancellation conditions. Only very few charge assignments are acceptable. 
A precise description of fermion masses and mixings is easily obtained by 
adjusting order one parameters. The sets of parameters that give a good
quantitative description of fermion data can be used to estimate the effects
of the fermion rotations on FCNC and CP violating processes
in supersymmetric models, with interesting prospects for $D^0$-$\bar D^0$
mixing, $B^0_s$-$\bar B^0_s$ mixing and $b \rightarrow s$ transitions.

\vskip0.3cm

\noindent {\bf Acknowledgments}\\
This work has been supported in part by the RTN European Program 
MRTN-CT-2004-503369. P.H.Ch. and S.P. were supported by the Polish State 
Committee for Scientific Research Grant 2 P03B 129 24 for 2003-2005.

\begin{appendix}

\section{Quark masses and CKM mixing angles}
\label{app:CKM}

In this appendix, we provide useful analytical expressions for quark
mass ratios and CKM angles in models with a spontaneously broken
horizontal abelian symmetry $U(1)_X$, at leading order in the small
parameter $\epsilon \equiv <\phi> / M$. We adopt the conventional
normalization $X_\phi = -1$, and assume that the $X$-charges of all
quark Yukawa couplings are positive, with $q_1 \geq q_2 \geq q_3$,
$\bar u_1 \geq \bar u_2 \geq \bar u_3$ and
$\bar d_1 \geq \bar d_2 \geq \bar d_3$. The Yukawa couplings
of the up and down quarks then read:
\begin{equation}
  Y^{AB}_u\ =\ C^{AB}_u\, \epsilon^{\bar u_A+q_B+h_u}\ ,  \quad
  Y^{AB}_d\ =\ C^{AB}_d\, \epsilon^{\bar d_A+q_B+h_d}\ ,
\end{equation}
where $C^{AB}_u$ and $C^{AB}_d$ are arbitrary parameters of order one.

We list below the expressions for the quark mass ratios $\frac{m_u}{m_t}$,
$\frac{m_c}{m_t}$, $\frac{m_d}{m_b}$ and $\frac{m_s}{m_b}$, and for the
CKM entries $V_{us}$, $V_{cb}$ and $V_{ub}$ at leading order in the
small parameter $\epsilon$. We  consider the three phenomenologically
relevant cases (see Section \ref{sec:constraints}):
(i) $\bar d_1 > \bar d_2 > \bar d_3$,
(ii) $\bar d_1 > \bar d_2 = \bar d_3$ and
(iii) $\bar d_1 = \bar d_2 > \bar d_3$, all with $q_1 > q_2 > q_3$
and $\bar u_1 > \bar u_2 > \bar u_3$.  In order to display compact
formulae, it is convenient to define the following combinations of
the coefficients $C^{AB}_u$:
\begin{eqnarray}
\Delta^{11}_u\ \equiv\ C^{22}_u C^{33}_u - C^{23}_u C^{32}_u\ ,
  & \Delta^{12}_u\ \equiv\ C^{21}_u C^{33}_u - C^{23}_u C^{31}_u\ ,
  & \Delta^{13}_u\ \equiv\ C^{21}_u C^{32}_u - C^{22}_u C^{31}_u\ , \qquad \\
  \Delta^{21}_u\ \equiv\ C^{12}_u C^{33}_u - C^{13}_u C^{32}_u\ ,
  & \Delta^{22}_u\ \equiv\ C^{11}_u C^{33}_u - C^{13}_u C^{31}_u\ ,
  & \Delta^{31}_u\ \equiv\ C^{12}_u C^{23}_u - C^{13}_u C^{22}_u\ , \qquad
\end{eqnarray}
and analogous combinations $\Delta^{AB}_d$ of the coefficients $C^{AB}_d$.
Note that $\det (C^{AB}_u) = C^{11}_u \Delta^{11}_u
- C^{12}_u \Delta^{12}_u +  C^{13}_u \Delta^{13}_u$, and similarly
$\det (C^{AB}_d) = C^{11}_d \Delta^{11}_d - C^{12}_d \Delta^{12}_d
+ C^{13}_d \Delta^{13}_d$.

\subsubsection*{Case (i): $q_1 > q_2 > q_3$,
$\bar u_1 > \bar u_2 > \bar u_3$ and $\bar d_1 > \bar d_2 > \bar d_3$}

In this case, it is convenient to define:
\begin{equation}
  \bar C_u^{AB}\ \equiv\ \frac{C_u^{AB}}{C_u^{33}}\ ,  \quad
  \bar C_d^{AB}\ \equiv\ \frac{C_d^{AB}}{C_d^{33}}\ . 
\end{equation}

The quark mass ratios are given by:
\begin{equation}
  \frac{m_u}{m_t}\ =\ \left| \bar C_u^{11}
    - \bar C_u^{12} \frac{\Delta_u^{12}}{\Delta_u^{11}}
    + \bar C_u^{13} \frac{\Delta_u^{13}}{\Delta_u^{11}} \right|
    \epsilon^{(q_1-q_3)+(\bar u_1-\bar u_3)}\ ,  \quad
  \frac{m_c}{m_t}\ =\ \frac{|\Delta_u^{11}|}{|C_u^{33}|^2}\,
    \epsilon^{(q_2-q_3)+(\bar u_2-\bar u_3)}\ ,
\end{equation}
\begin{equation}
  \frac{m_d}{m_b}\ =\ \left| \bar C_d^{11}
    - \bar C_d^{12} \frac{\Delta_d^{12}}{\Delta_d^{11}}
    + \bar C_d^{13} \frac{\Delta_d^{13}}{\Delta_d^{11}} \right|
    \epsilon^{(q_1-q_3)+(\bar d_1-\bar d_3)}\ ,  \quad
  \frac{m_s}{m_b}\ =\ \frac{|\Delta_d^{11}|}{|C_d^{33}|^2}\,
    \epsilon^{(q_2-q_3)+(\bar d_2-\bar d_3)}\ ,
\end{equation}
and the top and bottom Yukawa couplings are
$y_t = |C_u^{33}| \epsilon^{q_3+\bar u_3}$ and
$y_b = |C_d^{33}| \epsilon^{q_3+\bar d_3}$.

The CKM matrix entries are given by, in the standard phase convention
\cite{PDG04}:
\begin{eqnarray}
  V_{us} & = & \left| \frac{\Delta_d^{12}}{\Delta_d^{11}}
    - \frac{\Delta_u^{12}}{\Delta_u^{11}} \right|
    \epsilon^{q_1 - q_2}\ ,  \\
  V_{cb} & = & |\bar C_d^{32} - \bar C_u^{32}|\, \epsilon^{q_2 - q_3}\ ,  \\
  V_{ub} & = & \left| \bar C_d^{31}
    - \bar C_d^{32} \frac{\Delta_u^{12}}{\Delta_u^{11}}
    + \frac{\Delta_u^{13}}{\Delta_u^{11}}  \right|
    e^{-i \delta} \epsilon^{q_1 - q_3}\ ,  \\
  \delta  & = &  \arg \left( \frac{\bar C_d^{31}
    - \bar C_d^{32} \frac{\Delta_u^{12}}{\Delta_u^{11}}
    + \frac{\Delta_u^{13}}{\Delta_u^{11}}}
    {\left( \frac{\Delta_d^{12}}{\Delta_d^{11}}
    - \frac{\Delta_u^{12}}{\Delta_u^{11}} \right)
    (\bar C_d^{32} - \bar C_u^{32})}  \right)\ .
\end{eqnarray}

\subsubsection*{Case (ii): $q_1 > q_2 > q_3$,
$\bar u_1 > \bar u_2 > \bar u_3$ and $\bar d_1 > \bar d_2 = \bar d_3$}

In this case, it is convenient to define:
\begin{equation}
  \bar C_u^{AB}\ \equiv\ \frac{C_u^{AB}}{C_u^{33}}\ ,  \quad
  \bar C_d^{AB}\ \equiv\ \frac{C_d^{AB}}{\sqrt{|C_d^{23}|^2 +|C_d^{33}|^2}}\ . 
\end{equation}

The quark mass ratios are given by:
\begin{equation}
  \frac{m_u}{m_t}\ =\ \left| \bar C_u^{11}
    - \bar C_u^{12} \frac{\Delta_u^{12}}{\Delta_u^{11}}
    + \bar C_u^{13} \frac{\Delta_u^{13}}{\Delta_u^{11}} \right|
    \epsilon^{(q_1-q_3)+(\bar u_1-\bar u_3)}\ ,  \quad
  \frac{m_c}{m_t}\ =\ \frac{|\Delta_u^{11}|}{|C_u^{33}|^2}\,
    \epsilon^{(q_2-q_3)+(\bar u_2-\bar u_3)}\ ,
\end{equation}
\begin{equation}
  \hskip -.3cm \frac{m_d}{m_b}\ =\ \left| \bar C_d^{11}
    - \bar C_d^{12} \frac{\Delta_d^{12}}{\Delta_d^{11}}
    + \bar C_d^{13} \frac{\Delta_d^{13}}{\Delta_d^{11}} \right|
    \epsilon^{(q_1-q_3)+(\bar d_1-\bar d_3)} ,  \quad
  \frac{m_s}{m_b}\ =\ \frac{|\Delta_d^{11}|}{|C_d^{23}|^2 + |C_d^{33}|^2}\,
    \epsilon^{(q_2-q_3)+(\bar d_2-\bar d_3)} ,
\end{equation}
and the top and bottom Yukawa couplings are
$y_t = |C_u^{33}| \epsilon^{q_3+\bar u_3}$ and
$y_b = \sqrt{|C_d^{23}|^2 +|C_d^{33}|^2} \epsilon^{q_3+\bar d_3}$.

The CKM matrix entries are given by, in the standard phase convention:
\begin{eqnarray}
  V_{us} & = & \left| \frac{\Delta_d^{12}}{\Delta_d^{11}}
    - \frac{\Delta_u^{12}}{\Delta_u^{11}} \right|
    \epsilon^{q_1 - q_2}\ ,  \\
  V_{cb} & = & |\bar C_d^{22} \bar C_d^{23 \star}
    + \bar C_d^{32} \bar C_d^{33 \star}
    - \bar C_u^{32}|\, \epsilon^{q_2 - q_3}\ ,  \\
  V_{ub} & = & \left| (\bar C_d^{21} \bar C_d^{23 \star}
    + \bar C_d^{31} \bar C_d^{33 \star}) - (\bar C_d^{22} \bar C_d^{23 \star}
    + \bar C_d^{32} \bar C_d^{33 \star}) \frac{\Delta_u^{12}}{\Delta_u^{11}}
    + \frac{\Delta_u^{13}}{\Delta_u^{11}} \right|
    e^{-i \delta} \epsilon^{q_1 - q_3}\ ,  \\
  \delta  & = &  \arg \left( \frac{(\bar C_d^{21} \bar C_d^{23 \star}
    + \bar C_d^{31} \bar C_d^{33 \star})
    - (\bar C_d^{22} \bar C_d^{23 \star} + \bar C_d^{32} \bar C_d^{33 \star}) 
    \frac{\Delta_u^{12}}{\Delta_u^{11}} + \frac{\Delta_u^{13}}{\Delta_u^{11}}}
    {\left( \frac{\Delta_d^{12}}{\Delta_d^{11}}
    - \frac{\Delta_u^{12}}{\Delta_u^{11}} \right)
    (\bar C_d^{22} \bar C_d^{23 \star} + \bar C_d^{32} \bar C_d^{33 \star}
    - \bar C_u^{32})}  \right)\ .
\end{eqnarray}

\subsubsection*{Case (iii): $q_1 > q_2 > q_3$,
$\bar u_1 > \bar u_2 > \bar u_3$ and $\bar d_1 = \bar d_2 > \bar d_3$}

In this case, it is convenient to define:
\begin{equation}
  \bar C_u^{AB}\ \equiv\ \frac{C_u^{AB}}{C_u^{33}}\ ,  \quad
  \bar C_d^{AB}\ \equiv\ \frac{C_d^{AB}}{C_d^{33}}\ . 
\end{equation}

The quark mass ratios are given by:
\begin{equation}
  \frac{m_u}{m_t}\ =\ \left| \bar C_u^{11}
    - \bar C_u^{12} \frac{\Delta_u^{12}}{\Delta_u^{11}}
    + \bar C_u^{13} \frac{\Delta_u^{13}}{\Delta_u^{11}} \right|
    \epsilon^{(q_1-q_3)+(\bar u_1-\bar u_3)}\ ,  \quad
  \frac{m_c}{m_t}\ =\ \frac{|\Delta_u^{11}|}{|C_u^{33}|^2}\,
    \epsilon^{(q_2-q_3)+(\bar u_2-\bar u_3)}\ ,
\end{equation}
\begin{equation}
  \frac{m_d}{m_b}\ =\ \left| \bar C_d^{11}
    {\scriptstyle
    \frac{\Delta_d^{11}}{\sqrt{|\Delta_d^{11}|^2+|\Delta_d^{21}|^2}}}
    - \bar C_d^{21}
    {\scriptstyle
    \frac{\Delta_d^{21}}{\sqrt{|\Delta_d^{11}|^2+|\Delta_d^{21}|^2}}}
    + \bar C_d^{31}
    {\scriptstyle
    \frac{\Delta_d^{31}}{\sqrt{|\Delta_d^{11}|^2+|\Delta_d^{21}|^2}}} \right|
    \epsilon^{(q_1-q_3)+(\bar d_1-\bar d_3)}\ ,
\end{equation}
\begin{equation}
  \frac{m_s}{m_b}\ =\
    \frac{\sqrt{|\Delta_d^{11}|^2+|\Delta_d^{21}|^2}}{|C_d^{33}|^2}\,
    \epsilon^{(q_2-q_3)+(\bar d_2-\bar d_3)}\ ,
\end{equation}
and the top and bottom Yukawa couplings are
$y_t = |C_u^{33}| \epsilon^{q_3+\bar u_3}$ and 
$y_b = |C_d^{33}| \epsilon^{q_3+\bar d_3}$.

The CKM matrix entries are given by, in the standard phase convention:
\begin{eqnarray}
  V_{us} & = & \left| \frac{\Delta_d^{12} \Delta_d^{11 \star}
    + \Delta_d^{22} \Delta_d^{21 \star}}{|\Delta_d^{11}|^2+|\Delta_d^{21}|^2}
    - \frac{\Delta_u^{12}}{\Delta_u^{11}} \right| \epsilon^{q_1 - q_2}\ ,  \\
  V_{cb} & = & |\bar C_d^{32} - \bar C_u^{32}|\, \epsilon^{q_2 - q_3}\ ,  \\
  V_{ub} & = & \left| \bar C_d^{31}
    - \bar C_d^{32} \frac{\Delta_u^{12}}{\Delta_u^{11}}
    + \frac{\Delta_u^{13}}{\Delta_u^{11}}  \right|
    e^{-i \delta} \epsilon^{q_1 - q_3}\ ,  \\
  \delta  & = &  \arg \left( \frac{\bar C_d^{31} - \bar C_d^{32}
    \frac{\Delta_u^{12}}{\Delta_u^{11}} + \frac{\Delta_u^{13}}{\Delta_u^{11}}}
    {\left( \frac{\Delta_d^{12} \Delta_d^{11 \star}
    + \Delta_d^{22} \Delta_d^{21 \star}}{|\Delta_d^{11}|^2+|\Delta_d^{21}|^2}
    - \frac{\Delta_u^{12}}{\Delta_u^{11}} \right)
    (\bar C_d^{32} - \bar C_u^{32})}  \right)\ .
\end{eqnarray}

\section{Lepton masses and PMNS mixing angles}
\label{app:PMNS}

In this appendix, we provide analytical expressions for lepton masses and 
PMNS angles in models with a spontaneously broken horizontal abelian symmetry 
$U(1)_X$, at leading order in the small parameter 
$\epsilon\equiv\langle\phi\rangle/M$. We adopt the conventional normalization
$X_\theta = -1$, and assume that the $X$-charges of all lepton Yukawa 
couplings and right-handed neutrino mass terms are positive, with 
$l_1\geq l_2 \geq l_3$, $\bar n_1 \geq \bar n_2 \geq \bar n_3$ and 
$\bar e_1 \geq \bar e_2 \geq \bar e_3$. The Yukawa couplings of the charged 
leptons and the Dirac and Majorana matrices of the neutrinos then read:
\begin{equation}
Y^{AB}_e\ =\ C^{AB}_e \epsilon^{\bar e_A+l_B+h_d}\ ,
\quad Y^{AB}_D\ =\ C^{AB}_D \epsilon^{\bar n_A+l_B+h_u}\ , 
\quad M^{AB}_M\ =\ M_R C^{AB}_M \epsilon^{\bar n_A + \bar n_B}\ ,
\end{equation}
where $M_R$ is the scale of right-handed neutrino masses, and $C^{AB}_D$, 
$C^{AB}_e$ and $C^{AB}_M = C^{BA}_M$ are arbitrary parameters of order one. 
The seesaw mechism leads to the effective light neutrino mass matrix:
\begin{equation}
M^{AB}_\nu\ =\ C^{AB}_\nu \frac{v^2_u}{M_R}\, \epsilon^{l_A+l_B+2h_u}\ ,  
\quad
C^{AB}_\nu\ =\ C^{BA}_\nu\ =\ -\sum_{C,D} C^{CA}_D C^{DB}_D (C^{-1}_M)^{CD}\ ,
\label{eq:M_nu}    
\end{equation}
where we have used the fact that $(M^{-1}_M)^{CD} =
M^{-1}_R (C^{-1}_M)^{CD} \epsilon^{-\bar n_C - \bar n_D}$. It is a 
well-known fact that $M_\nu$ does not depend on the hierarchy of right-handed 
neutrinos (i.e. on the $\bar n_A$) when $\bar n_A+l_B+h_u \geq 0$ and 
$\bar n_A+\bar n_B \geq 0$, but only on the order one coefficients $C^{AB}_M$. 
It is then legitimate, from the low-energy point of view, to consider the 
$C^{AB}_\nu$ as arbitrary order one coefficients. As in the quark sector, we 
define combinations $\Delta^{AB}_e$ and $\Delta^{AB}_\nu$ of the order one 
coefficients $C^{AB}_e$ and $C^{AB}_\nu$ in order to display compact formulae 
for the lepton masses and mixings. For simplicity and given the fact that $CP$ 
violation has not been observed in the lepton sector yet, we shall assume that 
the coefficients $C^{AB}_e$ and $C^{AB}_\nu$ are real.

Given the structure (\ref{eq:M_nu}), neutrino data are best 
accommodated by $l_2 = l_3$ and $l_1 > l_3$, with the additional requirement 
that $x \equiv \frac{|\Delta^{11}_\nu|}{(C^{22}_\nu+C^{33}_\nu)^2} \ll 1$, 
to be explained by a (mild) cancellation among order one coefficients in 
$\Delta^{11}_\nu$, in order to account for the hierarchy between the solar 
and the atmospheric neutrino mass scales. In practice, only the cases 
$\epsilon^n\ll x\ll1$ and $\epsilon^n\sim x$, where $n\equiv l_1-l_3=1$ or 
$2$, can accommodate the LMA solution, and we shall discuss each of them in 
turn. Since the neutrino mass spectrum is hierarchical, the right-handed
neutrino mass scale has to be adjusted to
$M_R \sim v^2_u \epsilon^{2(l_3+h_u)} / \sqrt{\Delta m^2_{atm}}
\simeq (6 \times 10^{14}\, \mbox{GeV})\, \epsilon^{2(l_3+h_u)}$.
The mass hierarchy in the charged lepton sector requires 
$\bar e_1 > \bar e_2 > \bar e_3$ for $\epsilon \sim \lambda$, given that the 
acceptable values for $n$ are $1$ and $2$, as we shall see below.

The charged lepton mass ratios are given by:
\begin{equation}
\frac{m_e}{m_\tau}\ =\ \left| \bar C^{11}_e - \bar C^{12}_e 
\frac{\Delta^{12}_e}{\Delta^{11}_e}
+\bar C^{13}_e \frac{\Delta^{13}_e}{\Delta^{11}_e} \right|
\epsilon^{(l_1-l_3)+(\bar e_1 - \bar e_3)}\ ,
\quad  \frac{m_\mu}{m_\tau}\ =\
\frac{|\Delta^{11}_e|}{(C^{32}_e)^2+(C^{33}_e)^2}\,
\epsilon^{(\bar e_2 - \bar e_3)}\ ,
\end{equation}
where $\bar C^{AB}_e \equiv C^{AB}_e / \sqrt{(C^{32}_e)^2+(C^{33}_e)^2}$, 
and the tau Yukawa coupling is \\
$y_\tau = \sqrt{(C^{32}_e)^2+(C^{33}_e)^2}\, \epsilon^{l_3+\bar e_3+h_d}$.
The atmospheric and ``CHOOZ'' mixing angles $\theta_{23}$ and $\theta_{13}$ 
are given by:
\begin{eqnarray}
\tan\theta_{23} & = & \left| \frac{C^{33}_e C^{22}_\nu + C^{32}_e C^{23}_\nu}
{C^{32}_e C^{22}_\nu - C^{33}_e C^{23}_\nu} \right| ,  \\
\sin\theta_{13} & = & \left| 
\frac{(\Delta^{12}_e C^{22}_\nu + \Delta^{13}_e C^{23}_\nu)
(C^{22}_\nu + C^{33}_\nu) 
+ \Delta^{11}_e (C^{12}_\nu C^{22}_\nu + C^{13}_\nu C^{23}_\nu)}
{\Delta^{11}_e (C^{22}_\nu + C^{33}_\nu) 
\sqrt{(C^{23}_\nu)^2+(C^{33}_\nu)^2}} \right| \epsilon^n\ .
\end{eqnarray}
One naturally ends up with a large atmospheric mixing angle 
($\tan\theta_{23}\sim1$), as a result of the choice $l_2=l_3$, and with 
$\sin\theta_{13}\sim\epsilon^n$, which can be very close to the present 
experimental upper limit for $n=1$.

As for the neutrino masses and the solar mixing angle $\theta_{12}$, their 
formulae depend on whether $\epsilon^n \ll x$ or $\epsilon^n \sim x$.

\paragraph{Case 1: $\epsilon^n \ll x \ll 1$.}

The neutrino masses are given by, in units of $\frac{v^2_u}{M_R}$ and 
at leading order in $x$ and $\epsilon^n$:
\begin{eqnarray}
&& m_{\nu_1}\ =\ 
\frac{\left| C^{11}_\nu \Delta^{11}_\nu - C^{12}_\nu \Delta^{12}_\nu
+ C^{13}_\nu \Delta^{13}_\nu \right|}{(C^{22}_\nu+C^{33}_\nu)^2}\, 
\frac{\epsilon^{2n}}{x}\ ,  \\
& & m_{\nu_2}\ =\ |C^{22}_\nu + C^{33}_\nu|\, x\ ,  \\
& & m_{\nu_3}\ =\ |C^{22}_\nu + C^{33}_\nu|\, (1-x)\ ,
\end{eqnarray}
The ratio of solar to atmospheric neutrino oscillation frequencies is 
then approximately given by:
\begin{equation}
\frac{\Delta m^2_{21}}{\Delta m^2_{32}}\ \simeq\ x^2\ ,
\end{equation}
and the solar mixing angle reads:
\begin{equation}
\tan \theta_{12}\ =\ \left| \frac{\Delta^{12}_\nu}
{(C^{22}_\nu + C^{33}_\nu) 
\sqrt{(C^{23}_\nu)^2 + (C^{33}_\nu)^2}} \right| \frac{\epsilon^n}{x}\ .
\end{equation}
Neutrino oscillation data tell us that 
$\frac{\Delta m^2_{21}}{\Delta m^2_{32}} \approx 0.2$, hence $x\approx0.2$. 
For $n=2$ this is consistent with the present experimentally allowed range
for $\tan\theta_{12}$ ($0.28\leq\tan^2\theta_{12}\leq0.58$ at $3\sigma$ C.L.
\cite{BAGOPE}) provided that the order one coefficients conspire to give 
$\Delta^{12}_\nu /(C^{22}_\nu+C^{33}_\nu)\sqrt{(C^{23}_\nu)^2+(C^{33}_\nu)^2} 
\sim(2 - 3)$. The ``CHOOZ'' angle in this scheme is 
$\sin\theta_{13}\sim \epsilon^2 \simeq 0.05$, which corresponds to 
$\sin^2 2\theta_{13} \sim 0.01$, a value that should be accessible to the 
coming neutrino superbeam experiments T2K and NO\mbox{$\nu$}A.

\paragraph{Case 2: $\epsilon^n \sim x \ll 1$.}

The neutrino masses are given by, in units of $\frac{v^2_u}{M_R}$
and at leading order in $\epsilon^n \sim x$:
\begin{eqnarray}
&& m_{\nu_1}\ =\ |C^{22}_\nu+C^{33}_\nu|
\left|{\scriptstyle 1\, -\sqrt{1-\frac{4\det N}{(C^{22}_\nu+C^{33}_\nu)^3}\,
\frac{\epsilon^{2n}}{x^2}}} \right| \frac{x}{2}\ ,  \\
&& m_{\nu_2}\ =\ |C^{22}_\nu + C^{33}_\nu|
\left({\scriptstyle 1\, +\sqrt{1-\frac{4\det N}{(C^{22}_\nu + C^{33}_\nu)^3}\,
\frac{\epsilon^{2n}}{x^2}}} \right) \frac{x}{2}\ ,  \\
&& m_{\nu_3}\ =\ |C^{22}_\nu + C^{33}_\nu|\ .
\end{eqnarray}
The ratio of solar to atmospheric neutrino oscillation frequencies then reads:
\begin{equation}  
\frac{\Delta m^2_{21}}{\Delta m^2_{32}}\ =\ x^2\
\sqrt{1-\frac{4\det N}{(C^{22}_\nu+C^{33}_\nu)^3}\, 
\frac{\epsilon^{2n}}{x^2}}\ \sim\ x^2\ ,
\end{equation}
and the solar mixing angle is given by:
\begin{equation}
\tan^2 \theta_{12}\ =\ 
\frac{(\Delta^{12}_\nu \Delta^{13}_\nu)^2 + (C^{23}_\nu)^2
[(\Delta^{12}_\nu)^2 + (\Delta^{13}_\nu)^2] (m_{\nu_{1}} / \epsilon^n)^2}
{(\Delta^{12}_\nu \Delta^{13}_\nu)^2 + (C^{23}_\nu)^2
[(\Delta^{12}_\nu)^2 + (\Delta^{13}_\nu)^2] (m_{\nu_{2}} / \epsilon^n)^2}\ .
\end{equation}
Neutrino oscillation data require $x \sim 0.2$, hence $n=1$. The solar mixing 
angle naturally falls into the experimentally allowed range, and the 
``CHOOZ'' angle is predicted to be close to its present upper limit,
$\sin\theta_{13}\sim\epsilon \simeq 0.22$, which could be tested already
by the MINOS, CNGS (OPERA and ICARUS) or D-CHOOZ experiments.

\end{appendix}


\end{document}